\documentclass[a4paper]{revtex4}
\usepackage{mathrsfs}
\usepackage{graphicx}
\usepackage{latexsym}
\usepackage{amsmath}
\usepackage{amssymb}
\usepackage{textcomp}
\usepackage{amsbsy}
\usepackage{graphics}
\usepackage{color}
\usepackage{epstopdf}

\begin{document}
\title{\large \bf Scalar field collapse in Gauss-Bonnet gravity}
\author{Narayan Banerjee}
\email{narayan@iiserkol.ac.in}
\affiliation{Department of Physical Sciences, Indian Institute of Science Education and Research Kolkata,
Mohanpur Campus, Nadia, West Bengal 741246, India}
\author{Tanmoy Paul}
\email{pul.tnmy9@gmail.com}
\affiliation{Department of Theoretical Physics,\\
Indian Association for the Cultivation of Science,\\
2A $\&$ 2B Raja S.C. Mullick Road,\\
Kolkata - 700 032, India.\\}

\begin{abstract}
We consider a ``Scalar-Einstein-Gauss-Bonnet'' theory in four dimension, where the scalar field couples non minimally with 
the Gauss-Bonnet (GB) term. This coupling with the scalar field ensures the non topological character 
of the GB term. In this scenario, we examine the possibility for collapsing of the scalar field. Our result reveals that 
such a collapse is possible in the presence of Gauss-Bonnet gravity for suitable choices of parametric regions. 
The singularity formed as a result of the collapse is found to be a curvature singularity which is 
hidden from exterior by an apparent horizon.
\end{abstract}
\maketitle

\section{Introduction}
Over a few decades, relativistic astrophysics have gone through extensive developments, following the 
discovery of high energy phenomena in the universe such as gamma ray bursts. Compact objects like neutron stars have 
interesting physical properties, where the effect of strong gravity fields and hence general relativity are seen to play a 
fundamental role. A similar situation involving strong gravitational field, is that of a massive star undergoing a continual 
gravitational collapse at the end of its life cycle. This collapsing phenomena, dominated by the force of gravity, 
is fundamental in black hole physics and have received increasing attention in the past decades. The 
first systematic analysis of gravitational collapse in general relativity was given way back in 1939 by 
Oppenheimer and Snyder \cite{oppenheimer}. In this context, we also refer to the work by Datt\cite{datt}. Later developments 
in the study of gravitational collapse have been comprehensively summarized by Joshi\cite{joshi1,joshi2}.\\

Scalar fields have been of great interest in theories of gravity for various reasons. A scalar field 
with a variety of potential can fit superbly for cosmological requirements such as playing the role of the driver of the past 
or the present accelerated expansion of the universe. Apart from cosmological aspect, a suitable scalar potential can often mimic 
various matter distribution, including fluids with different equations of state. \\

Collapsing models with scalar fields are also quite well studied in General Relativity. The collapse of a zero mass scalar field 
was discussed by Christodoulou in \cite{1}. Consequently, the possibility of the formation of a naked singularity as 
an end product of a scalar field  collapse has also been explored in \cite{2}. Some variants of scalar field collapse and its 
implications are studied in \cite{4,5,6,7,8,9,10,17,18} (see also \cite{soumya,nb4,13,14,15,16}). \\

It is well known that Einstein-Hilbert action can be generalized by adding higher order curvature terms which naturally 
arise from diffeomorphism property of the action. Such terms also have their origin in String theory from quantum 
corrections. In this context F(R) \cite{24,25,26}, Gauss-Bonnet (GB) \cite{27,28,29} or more generally Lanczos-Lovelock 
gravity are some of the candidates in higher curvature gravitational theory. The spacetime curvature inside a collapsing 
star gradually increases as the collapse continues and becomes very large near the final state of the collapse. Thus for a 
collapsing geometry where the curvature becomes very large near the final state of the collapse, the higher curvature terms 
are expected to play a crucial role. Motivated by this idea, the collapsing scenarios in the presence of $F(R)$ gravity have 
been recently discussed by Goswami {\it et al}\cite{rituparno} and by Chakrabarti and Banerjee\cite{nb1,nb3}. \\

In the present context, we take the route of Gauss-Bonnet gravity to investigate the role of the higher order curvature in a 
scalar field collapse. The advantage of Gauss-Bonnet gravity is that the equations of motion do not contain any higher derivative 
terms (higher than two) of the metric, which lead to ghost free solution. The particular questions that we addressed in this paper 
are the following,

\begin{enumerate}
 \item Is the scalar field collapse possible in the presence of Gauss-bonnet gravity?
 
 \item If such a collapsing scenario is found, what is then the end product of the collapse, a black hole or a naked singularity?
\end{enumerate}

In order to address the above questions, we consider a ``Scalar-Einstein-Gauss-Bonnet'' theory \cite{nojiri_55} where 
the scalar field is coupled nonminimally to the GB term. The equations in a relativistic theory of gravity is already highly 
nonlinear and the presence of the Gauss-Bonnet terms make the situation even more difficult. In order to make the equations tractable, 
we start with a spatially homogeneous and isotropic model, i.e., essentially a Friedmann model. Except the scalar field we do not 
consider any other matter such as a fluid, but the collapse is homogeneous and thus somewhat analogous to the Oppenheimer-Snyder 
collapse\cite{oppenheimer}. \\

Our paper is organized as follows. In section II, we describe the model. In section III, we obtain the exact solution for the metric. 
Section IV and V address the visibility of the singularity produced as a result of the collapse and a matching of the solution with 
an exterior spacetime respectively. We end the paper with some concluding remarks in section VI.\\

\section{The model}
To explore the effect of Gauss-Bonnet gravity on scalar field collapse, we consider a ``Scalar-Einstein-Gauss-Bonnet'' theory 
in four dimension where the Gauss-Bonnet (GB) term is coupled 
with the scalar field. This coupling with the scalar field ensures the non topological character of the GB term. The 
action for this model 
\begin{eqnarray}
 S = \int d^4x \sqrt{-g} \bigg[R/(2\kappa^2)-(1/2)g^{\mu\nu}\partial_{\mu}\Phi\partial_{\nu}\Phi - V(\Phi)-\xi(\Phi)G\bigg],
 \label{action}\\
\end{eqnarray}
where $g$ is the determinant of the metric, $R$ is the Ricci scalar, $1/(2\kappa^2)=M_{p}^2$ is the four dimensional squared Planck scale 
and $G=R^2-4R_{\mu\nu}R^{\mu\nu}+R_{\mu\nu\alpha\beta}
R^{\mu\nu\alpha\beta}$ is the GB term, $\Phi$ denotes the scalar field also endowed with a potential $V(\Phi)$. 
The coupling between scalar field and GB term is symbolized by $\xi(\Phi)$ in the action. Variation of the action  
with respect to metric and scalar field leads to the field equations as follows,\\
\begin{eqnarray}
 &\frac{1}{\kappa^2}&[-R^{\mu\nu}+(1/2)g^{\mu\nu}R] + (1/2)\partial^{\mu}\Phi\partial^{\nu}\Phi - (1/4)g^{\mu\nu}\partial_{\rho}\Phi\partial^{\rho}\Phi + (1/2)g^{\mu\nu}\big[-V(\Phi)+\xi(\Phi)G\big]\nonumber\\ 
 &-&2\xi(\Phi)RR^{\mu\nu} - 4\xi(\Phi)R^{\mu}_{\rho}R^{\nu\rho} - 2\xi(\Phi)R^{\mu\rho\sigma\tau}R^{\nu}_{\rho\sigma\tau} + 4\xi(\Phi)R^{\mu\rho\nu\sigma}R_{\rho\sigma} + 2[\nabla^{\mu}\nabla^{\nu}\xi(\Phi)]R\nonumber\\ 
 &-&2g^{\mu\nu}[\nabla^2\xi(\Phi)]R - 4[\nabla_{\rho}\nabla^{\mu}\xi(\Phi)]R^{\nu\rho} - 4[\nabla_{\rho}\nabla^{\nu}\xi(\Phi)]R^{\mu\rho} + 4[\nabla^2\xi(\Phi)]R^{\mu\nu}\nonumber\\ 
 &+&4g^{\mu\nu}[\nabla_{\rho}\nabla_{\sigma}\xi(\Phi)]R^{\rho\sigma} + 4[\nabla_{\rho}\nabla_{\sigma}\xi(\Phi)]R^{\mu\rho\nu\sigma} = 0
 \label{gravitational equation}
 \end{eqnarray}
 and
 \begin{equation}
  g^{\mu\nu}[\nabla_{\mu}\nabla_{\nu}\Phi] - V'(\Phi) - \xi'(\Phi)G = 0,
  \label{scalar equation}
\end{equation}
where a prime denotes the derivative with respect to $\Phi$. It may be noticed that the gravitational equation of motion 
does not contain any derivative of the metric components of order higher than two. \\

The aim here is to construct a continual collapse model and for this purpose, we consider the following spherically 
symmetric non-static metric ansatz for the interior as, 
\begin{equation}
 ds^2 = -dt^2 + a^2(t)\big[dr^2 + r^2d\theta^2 + r^2\sin^2{\theta}d\varphi^2\big]
 \label{metric ansatz}
\end{equation}
where the factor $a(t)$ solely governs the interior spacetime characterized by the coordinates $t$, $r$, $\theta$ and $\varphi$. 
For such metric, the expression of Ricci scalar $R$ and GB term $G$ take the following form,
\begin{eqnarray}
 R = 6[2H^2 + \dot{H}]\nonumber\\
 G = 24H^2[H^2 + \dot{H}]
 \nonumber
\end{eqnarray}
with $H=\dot{a}/a$ and dot denotes the derivative with respect to time (t). Using the metric presented in eqn. (\ref{metric ansatz}), 
the field equations can be simplified and turn out to be,
\begin{equation}
 -(3/\kappa^2)H^2 + (1/2)\dot{\Phi}^2 + V(\Phi) + 24H^3\dot{\xi} = 0,
 \label{gr equation temporal part}
\end{equation}
\begin{eqnarray}
 \frac{1}{\kappa^2}\big[2\dot{H}+3H^2\big] + (1/2)\dot{\Phi}^2 - V(\Phi) - 8H^2\ddot{\xi} - 16H\dot{H}\dot{\xi} - 16H^3\dot{\xi} = 0,
 \label{gr equation spatial part}
\end{eqnarray}
and 
\begin{equation}
 \ddot{\Phi} +3H\dot{\Phi} + V'(\Phi) + 24\xi'(\Phi)(H^4+H^2\dot{H}) = 0.
 \label{scalar field equation}
\end{equation}
It is evident that due to the presence of GB term, cubic as well as quartic powers of the $H(t)$ appear in these equations.\\

It is well known that Einstein-Gauss-Bonnet gravity in 4-dimensions reduces to standard Einstein gravity, 
the additional terms actually cancel each other. In the present case, the nonminimal coupling with the scalar 
field assists the contribution from the GB term survive\cite{nojiri_55}. It is easy to see, in all the field 
equations above, that a constant $\xi$ (essentially no coupling) would immediately make the GB contribution trivial.

\section{Exact solutions: collapsing models}
In this section, we present a possible analytic solution of the field equations (eqn.(\ref{gr equation temporal part}), 
eqn.(\ref{gr equation spatial part}) and eqn.(\ref{scalar field equation})) and in order to do this, 
we consider a string inspired model \cite{nojiri_55} as follows,
\begin{eqnarray}
 V(\Phi) = V_0 e^{-2\Phi}
 \nonumber
\end{eqnarray}
and
\begin{eqnarray}
 \xi(\Phi) = \xi_0 e^{2\Phi},
\end{eqnarray}

where $V_0$ and $\xi_0$ are constants. Here we are interested on the collapsing solutions where the radius of the two sphere 
$ra(t)$ decreases monotonically with time. Using the above forms of $V(\Phi)$ and $\xi(\Phi)$, we 
solve the Friedmann equations for $a(t)$, $\Phi(t)$ and the solutions are following :
\begin{equation}
 a(t) = a_0 \big(t_0 -t\big)^{p},
 \label{sol_scale}
\end{equation}
and
\begin{equation}
 \Phi(t) = \ln {\big(t_0 - t\big)},
 \label{sol_scalar}
\end{equation}
where $a_0$, $t_0$ are constants of integration and the constant $p$ is related to $V_0$ and $\xi_0$ through the two relations, 
\begin{eqnarray}
 -3\frac{p^2}{\kappa^2} + \frac{1}{2} + V_0 + 48p^3\xi_0 = 0,
 \label{condition_1}\\
 -\frac{3p}{2} + \frac{1}{2} + V_0 - 24\xi_0p^3(p-1) = 0.
 \label{condition_2}
\end{eqnarray}
From the above relations, it can be easily shown that in absence of the coupling parameter $\xi_0$ (I.e., if $\xi_0=0$), the 
action reduces to the usual Einstein-Hilbert action and $p$ takes the value as $p=\frac{2}{3}$. Therefore, this particular 
value of the power exponent $p$ ($=\frac{2}{3}$) is not possible in the present context where the spacetime geometry evolves 
in the presence of ``Gauss-Bonnet'' term.  \\
The solution of $a(t)$ clearly reveals that $ra(t)$ decreases monotonically with time for $p > 0$. Therefore, the volume 
of the sphere of scalar field collapses with time and goes to zero at $t\rightarrow t_0$, giving rise 
to a finite time zero proper volume singularity. It is interesting to note that without the nonminimal coupling, the solution 
reduces to the time-reversed standard spatially flat Friedmann solution for a dust distribution (pressure $p=0$), here the 
solution is contracting rather than expanding. \\
In order to investigate whether the singularity is a curvature singularity or just an artifact of coordinate choice, one must 
look into the behaviour of Kretschmann curvature ($K$) scalar at $t\rightarrow t_0$. For the metric 
presented in eqn.(\ref{metric ansatz}), $K$ has the following expression,\\
\begin{equation}
 K = 6\bigg[\frac{\ddot{a}(t)^2}{a(t)^2} + \frac{\dot{a}(t)^4}{a(t)^4}\bigg]
 \label{curvature_scalar1}
\end{equation}

Using the solution of $a(t)$ (see eqn. (\ref{sol_scale})), the above expression of $K$ can be simplified as,\\
\begin{eqnarray}
 K = 6p^2\bigg[(p-1)^2 + p^2\bigg]\frac{1}{\big(t_0 - t\big)^4}
 \label{curvature_scalar2}
\end{eqnarray}

\begin{figure}[!h]
\begin{center}
 \centering
 \includegraphics[width=3.0in,height=2.0in]{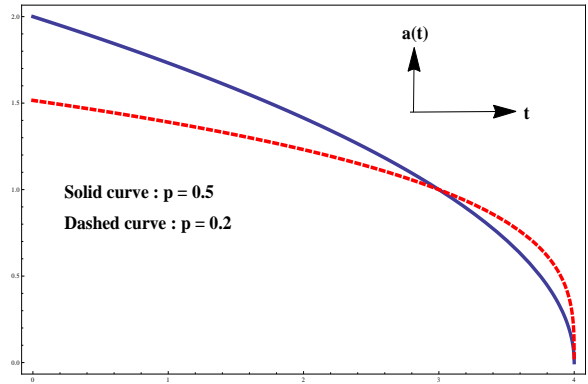}
 \caption{$a(t)$ vs $t$ for various values of $p$}
 \label{plot scale}
\end{center}
\end{figure}

It is clear from eqn.(\ref{curvature_scalar2}) that the Kretschmann scalar diverges at $t\rightarrow t_0$ and 
thus the collapsing sphere discussed here ends up in a curvature singularity.\\
From eqn. (\ref{sol_scale}), we obtain the plot (figure (\ref{plot scale})) between $a(t)$ and $t$ 
for two different values of $p$. 

Figure (\ref{plot scale}) clearly demonstrates that the spherical body collapses with time almost uniformly 
until $t$ approaches a value close to $t_0$, where it hurries towards a zero proper volume singularity.
This qualitative behaviour is almost not affected by different choices of $p$, only the rate of changes. 
We have presented here two specific examples with $p=0.5$ and $p=0.2$.\\

\section{Visibility of the singularity}
Whether the curvature singularity is visible to an exterior observer or not depends on the formation 
of an apparent horizon. The condition for such a surface is given by
\begin{eqnarray}
 g^{\mu\nu} Z_{,\mu} Z_{,\nu}\bigg|_{r_{ah},t_{ah}} = 0
 \label{app hor 1}
\end{eqnarray}

where $Z$ is the proper radius of the two sphere, given by $ra(t)$ in the present case, $r_{ah}$ and 
$t_{ah}$ being the comoving radial coordinate and time of formation of the apparent horizon respectively. 
Using the form of $g^{\mu\nu}$ presented in eqn.(\ref{metric ansatz}), above expression can be simplified 
and turns out to be,
\begin{eqnarray}
 r_{ah}^2 \dot{a}(t_{ah})^2 = 1
 \label{app hor 2}
\end{eqnarray}

The solution of $a(t)$ (see eqn.(\ref{sol_scale})) immediately leads to eqn.(\ref{app hor 2}) as follows :
\begin{eqnarray}
 \big[t_0 - t_{ah}\big]^{2p-2} = \frac{1}{(r_{ah}^2 a_0^2 p^2)}
 \label{app hor 3}
\end{eqnarray}
The above equation clearly reveals that $t_{ah}$ is less than $t_0$ (i.e. $t_{ah} < t_0$). 
Therefore the apparent horizon forms before the formation of singularity. Thus, the curvature singularity 
discussed here is always covered from an exterior observer by the formation of an 
apparent horizon. It may be mentioned that, the singularity is independent of the radial coordinate $r$ and it is covered by 
a horizon. This result is consistent with the result obtained by Joshi {\it et al} \cite{31_NS} that unless one has a 
central singularity, it can not be a naked singularity. \\

\section{Matching of the interior spacetime with an exterior geometry}

To complete the model, the interior spacetime geometry of the collapsing sphere needs to be matched to 
an exterior geometry. For the required matching, the Israel conditions are used, 
where the metric coefficients and extrinsic curvatures (first and second fundamental forms 
respectively) are matched at the boundary of the sphere. Following references \cite{18,66}, we match 
the interior spacetime with a generalized Vaidya exterior spacetime at the boundary hypersurface $\Sigma$ 
given by $r = r_0$. The metric inside and outside of $\Sigma$ are given by,
\begin{eqnarray}
 ds_{-}^2 = -dt^2 + a^2(t)\bigg[dr^2 + r^2d\theta^2 + r^2\sin^2{\theta}d\varphi^2\bigg]
 \label{inside metric}
\end{eqnarray}
and
\begin{eqnarray}
 ds_{+}^2 = -\bigg(1 - \frac{2M(r_v,v)}{r_v}\bigg)dv^2 + 2dvdr_v + r^2d\theta^2 + r^2\sin^2{\theta}d\varphi^2
 \label{outside metric}
\end{eqnarray}
respectively, where $r_v,v,\theta$ and $\varphi$ are the exterior coordinates and $M(r_v,v)$ is known 
as generalized mass function. The same hypersurface $\Sigma$ can alternatively be defined by the exterior 
coordinates as $r_v = R(t)$ and $v = T(t)$. Then the metrics on $\Sigma$ from inside and 
outside coordinates turn out to be,
\begin{eqnarray}
 ds_{-,\Sigma}^2 = -dt^2 + a^2(t)r_0^2 d\Omega^2
 \nonumber
\end{eqnarray}
and
\begin{eqnarray}
 ds_{+,\Sigma}^2 = -\bigg[\bigg(1 - \frac{2M_{\Sigma}(t)}{R(t)}\bigg)\dot{T}^2 - 2\dot{T}\dot{R}\bigg]dt^2 + R(t)^2 d\Omega^2
 \nonumber
\end{eqnarray}
where $M_{\Sigma}(t)$ ($=M(R(t),T(t))$) is the generalized mass function on $\Sigma$, $d\Omega^2$ denotes the line 
element on a unit two sphere and dot represents $\frac{d}{dt}$. Matching the first fundamental 
form on $\Sigma$ (i.e. $ds_{-,\Sigma}^2 = ds_{+,\Sigma}^2$) yields the following two conditions :
\begin{eqnarray}
 \frac{dT(t)}{dt} = \frac{1}{\sqrt{1 - \frac{2M_{\Sigma}(t)}{R(t)} - 2\frac{dR(t)}{dT(t)}}}
 \label{con 1}
\end{eqnarray}
and
\begin{eqnarray}
 R(t)&=&r_0a(t)\nonumber\\
 &=&r_0a_0 (t_0 - t)^{p}
 \label{con 2}
\end{eqnarray}
In order to match the second fundamental form, we calculate the normal of the hypersurface $\Sigma$ 
from inside ($\vec{n}_{-} = n_{-}^t$, $n_{-}^r$, $n_{-}^{\theta}$, $n_{-}^{\varphi}$) and outside 
($\vec{n}_{+} = n_{+}^v$, $n_{+}^{r_v}$, $n_{+}^{\theta}$, $n_{+}^{\varphi}$) coordinates as follows :
\begin{eqnarray}
 &n&_{-}^t = 0~~,~~~~~~~~n_{-}^r =  a(t)~~,~~~~~~~n_{-}^{\theta} = n_{-}^{\varphi} = 0\label{inside normal}
\end{eqnarray}
and
\begin{eqnarray}
 n_{+}^v = \frac{1}{\sqrt{1 - \frac{2M_{\Sigma}(t)}{R(t)} - 2\frac{dR(t)}{dT(t)}}}~~,\nonumber
 \end{eqnarray}
 \begin{eqnarray}
 n_{+}^{r_v} = \frac{1 - \frac{2M_{\Sigma}(t)}{R(t)} - \frac{dR(t)}{dT(t)}}
 {\sqrt{1 - \frac{2M_{\Sigma}(t)}{R(t)} - 2\frac{dR(t)}{dT(t)}}}~~,\nonumber
 \end{eqnarray}
 \begin{eqnarray}
 n_{+}^{\theta} = n_{+}^{\varphi} = 0.
\label{outside normal}
\end{eqnarray}

The above expressions of $\vec{n}_{-}$ and $\vec{n}_{+}$ leads to the extrinsic 
curvature of $\Sigma$ from interior and exterior coordinates respectively, and are given by,
\begin{eqnarray}
 K_{tt}^- = 0~~,~~~~~~~~~K_{\theta\theta}^- = r_0a(t)~~,~~~~~~~K_{\varphi\varphi}^- = r_0a(t)\sin^2{\theta}
 \label{inside extrinsic}
\end{eqnarray}
from interior metric, and
\begin{eqnarray}
 K_{tt}^+ = \frac{\partial M_{\Sigma}(t)}{\partial R(t)} - \frac{M_{\Sigma}(t)}{r_0a(t)} - r_0^2 a(t)\ddot{a}(t)~~,
 \nonumber
\end{eqnarray}
\begin{eqnarray}
 K_{\theta\theta}^+ =  R(t) \frac{1 - \frac{2M_{\Sigma}(t)}{R(t)} - \frac{dR(t)}{dT(t)}}
 {\sqrt{1 - \frac{2M_{\Sigma}(t)}{R(t)} - 2\frac{dR(t)}{dT(t)}}}~~,
 \nonumber
\end{eqnarray}
\begin{eqnarray}
 K_{\varphi\varphi}^+ = R(t)\sin{\theta}^2 \frac{1 - \frac{2M_{\Sigma}(t)}{R(t)} - \frac{dR(t)}{dT(t)}}
 {\sqrt{1 - \frac{2M_{\Sigma}(t)}{R(t)} - 2\frac{dR(t)}{dT(t)}}}
 \label{outside extrinsic}
\end{eqnarray}
from exterior metric.\\
The equality of the extrinsic curvatures of $\Sigma$ from both sides is therefore equivalent to the following 
two conditions :
\begin{eqnarray}
 r_0 a(t) = R(t) \frac{1 - \frac{2M_{\Sigma}(t)}{R(t)} - \frac{dR(t)}{dT(t)}}
 {\sqrt{1 - \frac{2M_{\Sigma}(t)}{R(t)} - 2\frac{dR(t)}{dT(t)}}}
 \label{con 3}
\end{eqnarray}
and
\begin{eqnarray}
 \frac{\partial M_{\Sigma}(t)}{\partial R(t)} = \frac{M_{\Sigma}(t)}{r_0a(t)} + r_0^2 a(t)\ddot{a}(t)
 \label{con 4}
\end{eqnarray}

By using eqn. (\ref{con 1}), eqn. (\ref{con 2}) and eqn. (\ref{con 3}), the mass 
function on $\Sigma$ (i.e. $M_{\Sigma}(t)$) can 
be determined and is given by,
\begin{eqnarray}
 M_{\Sigma}(t)&=&\frac{1}{2} R(t)\dot{R}^2(t)\nonumber\\
 &\propto&(t_0 - t)^{3p-2}\nonumber\\
 &\propto&\rho a^3(t)
 \label{mass}
\end{eqnarray}
where $\rho$ ($= \frac{1}{2}\dot{\Phi}^2 + V(\Phi)$) is energy density of the scalar field $\Phi$. 
Moreover with eqn. (\ref{con 1}) 
and eqn.(\ref{mass}), one finally lands with the following expression,
\begin{eqnarray}
 \frac{dT(t)}{dt} = \frac{1}{\big(1 + r_0a_0 p(t_0 - t)^{p-1}\big)}.
 \label{con 5}
\end{eqnarray}

Eqn.(\ref{con 2}), eqn.(\ref{con 4}), eqn.(\ref{mass}) and eqn.(\ref{con 5}) completely 
specify the matching at the boundary of the collapsing scalar field with an exterior generalized Vaidya geometry. 
However all the matching conditions are not independent, but eqn.(\ref{con 4}) 
can be derived from the other three conditions.\\
It deserves mention that, if the interior spacetime is matched with exterior ``Schwarzschild geometry'', 
then the only possible value that can be acquired by the parameter '$p$' is $p = \frac{2}{3}$. But 
as discussed earlier, that for this particular value of $p$ ($=\frac{2}{3}$), the Gauss-Bonnet coupling parameter ($\xi_0$) 
goes to zero which in turn annihilates the GB term from the action. Thus in the present context, the presence of Gauss-Bonnet 
gravity spoils the matching of interior geometry of the collapsing cloud with an exterior 
Schwarzschild geometry. This is quite consistent because the presence of Gauss-Bonnet term generates an effective 
energy momentum tensor which can not be zero (since it arises effectively from spacetime curvature) at the outside of $\Sigma$ and hence 
no exterior vacuum like Schwarzschild solution is matched with the interior spacetime metric in the presence of Gauss-Bonnet term.\\

Before concluding, it may be mentioned that the positivity of energy density requires $\frac{\partial M_{\Sigma}(t)}{\partial t} > 0$. 
This condition along with eqn.(\ref{mass}) constraints the parameter $p$ as less than $\frac{2}{3}$ (i.e. $p < \frac{2}{3}$).

\section{Conclusion}
We consider a ``Scalar-Einstein-Gauss-Bonnet'' theory in four dimensions where the scalar field 
couples non minimally with the Gauss-Bonnet term. This coupling with the scalar field ensures the 
non topological character of the GB term. In this scenario, we examine the possibility of collapse
of the scalar field by considering a spherically symmetric spacetime metric. \\

With the aforementioned metric, an exact solution is obtained for the spacetime geometry, which clearly 
reveals that the radius of a two sphere decreases monotonically with time if the parameter $p$ is taken 
to be greater than zero. This parameter $p$ is actually determined by the strength of the coupling of the scalar 
field to the GB term and a parameter ($V_0$) by the relations (\ref{condition_1}) 
and (\ref{condition_2}). \\

From the behaviour of Kretschmann scalar, it is found that the singularity formed as a result of the collapse is a finite 
time curvature singularity. Moreover the scalar field energy density also seems to be divergent at the singularity. The 
formation of apparent horizon is investigated and it turns out that the apparent horizon forms before the formation of singularity. 
Therefore the curvature singularity discussed here is hidden from exterior by an apparent horizon.\\

Finally, we match the interior spacetime geometry of the collapsing sphere with generalized Vaidya exterior geometry at 
the boundary of the cloud ($\Sigma$). For this matching, the Israel junction conditions are used where the metric 
coefficients and extrinsic curvatures are matched on $\Sigma$. We determine the matching conditions given 
in eqn.(\ref{con 2}), eqn.(\ref{con 4}), eqn.(\ref{mass}) and eqn.(\ref{con 5}). We also investigate whether the 
interior geometry can be matched with a Schwarzschild exterior geometry or not. It is found that the interior spacetime 
is matched with the exterior Schwarzschild for the only possible value of '$p$' as $p = \frac{2}{3}$. But for this 
particular value of $p$ ($=\frac{2}{3}$), the Gauss-Bonnet coupling parameter ($\xi_0$) goes to zero which in turn vanishes 
the GB term from the action. Thus in the present context, the presence of Gauss-Bonnet gravity spoils the matching of interior 
geometry of the collapsing cloud with an exterior 
Schwarzschild geometry. This result is in fact quite consistent because the presence of Gauss-Bonnet term generates an effective 
energy momentum tensor which can not be zero (since it arises effectively from spacetime curvature) at the outside of $\Sigma$ and hence 
no exterior vacuum like Schwarzschild solution is matched with the interior spacetime metric in the presence of Gauss-Bonnet term.\\

Another important point to be mentioned that the singularity formed is not a central singularity, it is formed at any value of $r$ within 
the distribution. Such a singularity general relativity is always covered by a horizon\cite{31_NS}. It is interesting to note that the result 
obtained in the present work in the presence of Gauss-Bonnet term is completely consistent with the corresponding GR result.

\end{document}